\documentclass[useAMS,usenatbib]{mnras}

\usepackage{bm}
\usepackage{natbib}
\usepackage{graphicx}
\usepackage{amssymb, amsmath}
\usepackage[normalem]{ulem}
\usepackage[dvipsnames]{xcolor}

\title[Composition of the upper layers of the inner crust]{Accreting neutron stars: composition of the upper layers of the inner crust}
\author[N.\ N.\ Shchechilin, M.\ E.\ Gusakov, A.\ I.\ Chugunov]
{N.\ N.\ Shchechilin,$^{1,}$\thanks{nikolai.shchechilin@ulb.be} M.\ E.\ Gusakov,$^{2}$ A.\ I.\ 
Chugunov$^{2}$\\
$^{1}$Institut d'Astronomie et d'Astrophysique, CP-226, Universit\'e Libre de Bruxelles, 1050, Brussels, Belgium\\	
$^{2}$Ioffe Institute, Polytekhnicheskaya 26, 194021 Saint Petersburg, Russia}
\begin{document}
	
	\date{Accepted 2022 xxxx. Received 2022 xxxx;		in original form 2022 xxxx}
	
	\pagerange{\pageref{firstpage}--\pageref{lastpage}}
	\pubyear{2022}
	
	\maketitle
	
	\label{firstpage}

%%%%%%%%%%%%%%%%%%%%%%%%%%%%%%%%%%%%%%%%%%%%%%%%%%%%%%%%%%%%%%%%%%%%%%%%%%%%%%
\begin{abstract}
We model the nuclear evolution of an accreted matter 
as it sinks toward the stellar center, in order to find its composition  
and equation of state.
To this aim, we developed
a simplified reaction network 
that allows
for redistribution of free neutrons in the inner crust to
satisfy the recently suggested neutron hydrostatic and diffusion equilibrium condition. 
We
analyse
the 
main
reaction pathways for the three representative thermonuclear ash compositions: Superburst, Kepler,  
and Extreme rp.
In contrast to the previous results, which neglect redistribution of free (unbound) neutrons in the 
inner crust, the most 
significant 
reactions 
in our calculations
are neutron captures and electron emissions. 
The pycnonuclear fusion plays 
some
role only for Kepler ashes.
For the direct application of our results in astrophysical codes we present profiles of the average charge, $\langle Z\rangle$, impurity parameter, $Q_\mathrm{imp}$ and equation of state for a set of models,  parametrized by the pressure at the outer-inner crust interface. 
Typically, for Superburst ashes
$Q_\mathrm{imp}\approx 1-4$,
while for Kepler ashes  $Q_\mathrm{imp}$ decreases from $\approx23$ at the outer-inner crust 
interface 
to $\approx5$ at the end of our simulation
(the corresponding density equals $\rho_\mathrm{dc}\approx2\times 
10^{12}$~g\,cm$^{-3}$).
At the same time,
for Extreme rp ashes $Q_\mathrm{imp}$ remains large $\approx 30-35$ in the considered inner crust 
region.
Our results 
are important 
for modeling 
the thermal relaxation of
transiently accreting neutron stars after the end of the outburst.   
\end{abstract}
%%%%%%%%%%%%%%%%%%%%%%%%%%%%%%%%%%%%%%%%%%%%%%%%%%%%%%%%%%%%%%%%%%%%%%%%%%%%%%


%%%%%%%%%%%%%%%%%%%%%%%%%%%%%%%%%%%%%%%%%%%%%%%%%%%%%%%%%%%%%%%%%%%%%%%%%%%%%%
\section{Introduction}
%%%%%%%%%%%%%%%%%%%%%%%%%%%%%%%%%%%%%%%%%%%%%%%%%%%%%%%%%%%%%%%%%%%%%%%%%%%%%%

Accreting neutron stars (NSs) in low-mass X-ray binaries (LMXBs) 
provide intriguing opportunity to observe NS response to the perturbation, caused by accretion. Namely, in such systems the matter from the low-mass ($M\lesssim M_\odot$) donor star is transferred by Roche-lobe overflow 
via
the accretion disc 
to the NS surface. 
It is generally assumed that accreted matter spreads rapidly across the surface of the NS due to 
rather modest NS magnetic field ($B\sim 10^8-10^{10}$ G, see, e.g., \citealt{wdp17}). 
The newly
accreted material compresses the NS crust and triggers nuclear reactions there,
which lead to crust heating 
(the so called deep crustal heating paradigm, 
\citealt*{HZ90b,bbr98,Bildsten+01}). 
The accretion process is 
non-stationary
due to disc instabilities (\citealt{Lasota01}), 
so that
some of the accreting NSs 
were observed 
in quiescent state, when the accretion is halted, and thermal emission from the NS surface 
was
detected 
or constrained
(see \citealt*{pcc19} for a compilation of observational results and the corresponding references).
Analysis of these observations and comparison with theoretical models
can 
give us valuable
information about the properties of superdense matter in NS interiors (see \citealt{mdkse18} for a 
review).

According to the generally accepted theoretical concept, 
thermal evolution of transiently accreting neutron stars is associated with numerous cycles, consisting of the
periods of 
accretion interspersed by quiescence intervals. During such cycles,  the heat generated by nuclear 
reactions in the crust partially flows inward, to the NS core, and partially 
accumulates
in the 
overheated 
crust. 
The latter heat can lead to
observable thermal relaxation of the crust after accretion 
stops.
If another accretion episode does not occur, the crust reaches thermal equilibrium with the 
core, arriving at a fairly stable thermal configuration, where long-term heating and cooling are 
balanced appropriately.

In line with
this  
concept, 
theoretical efforts 
have been focused 
on the two research directions.
Historically, the first direction was aimed at describing the thermal equilibrium state and 
analyzing the
dependence of the thermal luminosity in quiescence, $L$, on the average mass accretion rate, 
$\dot{M}$.
In this case,
theoretical $L(\dot{M})$ diagrams were constructed and confronted with 
observational data on LMXBs 
(\citealt*{bbr98,ylh03,ylpgc04,yp04,hjwt07,heinke_et_al_10,wdp13,by15,hs17,pcc19, 
Fortin_ea21,Mendes_ea22}). The $L(\dot{M})$ diagrams are most sensitive to the physics of the core 
(primarily via the neutrino emissivity and heat capacity), thus properties of the 
core had been constrained by this method.

The second 
research direction 
is mainly associated with the analysis of the
crust thermal relaxation, 
which 
was 
observed for several accreting NSs just after the end of accretion episode.
Theoretical modeling of 
this process 
depends mainly on the crust 
properties, which 
can thus be
constrained 
by the analysis of 
crust cooling observations
(\citealt{syhp07,bc09,pr12,wdp17,mdkse18,pc21};
see, however, \citealt{Brown_ea18,Mendes_ea22} for the constraints on the physics of the core).
In particular, \cite{syhp07} revealed the evidences of relatively high heat conductivity in the 
crust,
suggesting that the latter   
solidifies in a rather pure crystal with insignificant amount of impurities.
The most intriguing result 
of subsequent works
was the
conclusion
that deep crustal heating 
is
insufficient for explaining the
thermal relaxation of the crust
for 
many 
observed sources.
In order to explain the observations, a number of authors were forced to include
phenomenological shallow heating sources in their simulations
(\citealt{bc09, Deibel+15,Waterhouseetal16,Parikh+18,Parikh_ea19,Page+22,LuLuLiu+22}). 
Unfortunately, 
the nature of these sources 
remains largely unidentified. 

For both research methods described above 
it is crucial to know the amount of deep crustal heating and the crust composition 
in order to have a reliable theoretical basis for 
modeling the thermal evolution.         
These parameters 
depend on the initial composition of the outermost layers of the crust, which are pushed 
toward the stellar center
during accretion. 
The initial composition is formed in the thermonuclear burning of accreted elements in the NS 
envelope. 
The burning proceeds in different regimes (stable or unstable)
depending
on the composition of 
the accreted matter (the amount of H, He, and metallicity) and the accretion rate 
(\citealt{Strohmayer&Bildsten06,johnston20,Galloway+21}). As a result, the composition of 
thermonuclear burning ashes varies in different models. In particular,  \cite{Cyburt_ea16}, using 
1D multi-zone model based on the hydrodynamic code Kepler  
(\citealt{Weaver_ea78_Kepler,Woosley_ea04_KeplerCode}),  found a wide distribution of elements with 
the most abundant nuclides being in the iron group.
However, if a large enough amount of hydrogen is preserved at the moment of burst ignition, the 
extended rapid proton capture process could lead to the formation of the palladium group elements 
(Extreme rp ash, \citealt{Schatz_ea01}). In addition, if under certain circumstances sufficient 
amount of carbon was accumulated, a superburst could occur (see, e.g., 
\citealt{Dohi_ea22_superburst,Meisel22_Superburst} for recent studies), leading to the narrow 
distribution of 
nuclides
with a peak 
in the elements of the iron group
(Superburst ashes, 
\citealt{KH11}). 
Because of many alternatives,
it is important
to be able to predict
the properties of 
the accreted crust matter for different
models of thermonuclear ashes.

The traditional approach for calculating the nuclear evolution of the thermonuclear burning ashes was pioneered by
\cite{Sato79}.
In this approach one generally follows the
nuclear reactions during compression of a matter element with a fixed number of baryons. The most 
widely applied is the one-component model of \cite{HZ90,HZ03,HZ08} 
(see \citealt{Fantina_ea18,Fantina+22} for a recent update 
employing
the modern energy-density functional to describe nuclear microphysics). 
Multicomponent models (\citealt{gkm08,Steiner12,lau_ea18,SC19_MNRAS,Schatz_ea22}) 
deal with 
a much richer reaction network and, as a consequence, allow for
a more detailed investigation of the crustal 
properties.
Within the traditional approach the general physics of the accreted crust
appeared to be well-established.
Namely, it was 
found
that the main reactions in the crust are electron captures, neutron emissions and pycnonuclear 
fusions, and the deep crustal heating produces $(1.5-2)$\,MeV per accreted baryon (although this value 
can be effectively reduced by neutrino emission associated with electron 
capture/emission cycles, see \citealt{Schatz_Nat14,lau_ea18,Schatz_ea22}). 
However, \citealt{CS19_NoEquil} pointed out that the traditional approach 
has a serious shortcoming: it
neglects redistribution of free (unbound) neutrons in the inner crust.
Thus, there is a need for reconsideration of all the previous results.

As shown by \cite{GC20_DiffEq} (GC20), 
neutrons rapidly redistribute themselves 
in
the inner crust in order to stay 
in hydrostatic and diffusion equilibrium, in which the redshifted neutron chemical potential 
$\mu_\mathrm{n}^{\infty}$ 
remains constant throughout the inner crust 
(below this condition is 
abbreviated as
the nHD-condition, 
from `neutron 
Hydrostatic and Diffusion equilibrium').
GC20 demonstrated that accretion process leads to formation of the fully accreted crust (FAC), 
which 
has stationary structure.%
\footnote{Strictly speaking, this statement holds true only if the composition of thermonuclear 
ashes does not evolve in time. Generally, it is not the case and different 
burning regimes
can 
operate
in the shallow regions of the same NS (see, e.g., 
\citealt{Li_ea21_Unstable_Stable_Transition} for the recent observational evidences). However, we 
do 
not consider this possibility here and only construct models
with a 
fixed composition of thermonuclear ashes.
}
Crucial element of the FAC formation is the mechanism of nuclei disintegration, which prevents 
accumulation of nuclei in the crust.
GC20 revealed 
that this mechanism naturally arises 
in
the nHD approach and illustrated the results 
within the compressible liquid drop model based on the SLy4 energy-density functional 
(\citealt{Chabanat_ea98_SLY4,DHM00}).
Later, \cite{GC21_HeatReleaze}  (GC21)  derived a general formula for the total heat release, which 
allows one to determine the inner crust heating without a detailed 
knowledge
of the crust 
structure. GC21 employed more 
realistic nuclear models (accounting for shell effects) to estimate the heating and obtained 
$\approx0.5$\,MeV per baryon, which 
is a factor of few lower than in the traditional models. 
However, numerical results of GC21 
were obtained for a 
one-component 
model of the crust matter.

These results were further elaborated in \cite*{SGC_OC21} 
(SGC21 in what follows), 
where we 
considered realistic multicomponent ashes. 
Applying a simplified reaction network, we calculated 
nuclear evolution in the outer crust and, using the general heating formula from GC21,  
we obtained the total heat release in the whole crust.
For Kepler and Superburst ashes, dominated by the iron-group nuclei, the results well agree with 
the pure $^{56}$Fe ash
employed
in GC21. In contrast, for Extreme rp ash, dominated by the 
palladium-group nuclei, the deep crustal heating can be even lower, 
$\approx0.2$\,MeV per 
baryon.

Later, \citealt*{GKC_psi21} (GKC21) analysed thermodynamics of nHD inner crust and demonstrated, 
that the appropriate thermodynamic potential that should be minimized 
there
is $\Psi=G-\mu_\mathrm{n} N_\mathrm{b}$ (where $G$ is the usual Gibbs potential and $N_\mathrm{b}$
is the number of baryons in a given matter element). 
Further we (\citealt*{SGC22a}, SGC22) made use of this potential to directly calculate the heat 
release in the upper layers of the inner crust for three thermonuclear ashes and thus extend our 
previous 
work
SGC21. Here we present details of the simulations performed in SGC22:  nuclear reaction pathways 
and the composition in the upper layers of the inner crust.

In section~\ref{Sec:Approach} we briefly describe 
the main ideas behind our calculations.
In section~\ref{Sec:res} we demonstrate the major reaction pathways. In contrast to the traditional 
models, the main reactions are neutron captures and electrons emissions.
We also present profiles of the average charge number $\langle Z\rangle$ and impurity parameter 
$Q_\mathrm{imp}$, in addition to showing the crustal equation of state. 
We outline our main conclusions in the section~\ref{Sec:sum}.  

%%%%%%%%%%%%%%%%%%%%%%%%%%%%%%%%%%%%%%%%%%%%%%%%%%%%%%%%%%%%%%%%%%%%%%%%%%%%%%
\section{Basic features of the nHD approach and simplified reaction network}
\label{Sec:Approach}
%%%%%%%%%%%%%%%%%%%%%%%%%%%%%%%%%%%%%%%%%%%%%%%%%%%%%%%%%%%%%%%%%%%%%%%%%%%%%%

In the present paper we supplement the energy release calculations of SGC22 with the details of 
nuclear evolution and composition profiles. 
For the sake of completeness, below we summarize the principal ideas of the
nHD 
approach, used in the SGC22, and present a brief overview of our previous results.

The nHD approach fundamentally differs from the traditional method of studying the accreted crust.
First, it is based on a physically consistent treatment of neutrons in the inner crust. Specifically, neutrons 
are allowed to move and redistribute 
freely between
adjacent
crustal layers by 
means of 
superfluid 
currents and
diffusion. 
As shown in GC20, in both these cases the neutron redistribution leads to the nHD condition, 
$\mu_\mathrm{n}^{\infty}=\mu_\mathrm{n} e^{\nu/2}=\rm{const}$, where $e^{\nu/2}$ is the 
redshift factor. 
As a result, the neutron number density in the inner crust varies smoothly.

One can think about the inner crust as consisting of a `neutron sea'
of
unbound neutrons 
and 
nuclei, embedded into this sea. During the accretion process nuclei sink in the sea, being 
pushed to the deeper layers of the crust.
By definition, the sea surface corresponds to the interface between the outer and inner crust 
(oi interfcace),
the pressure at the interface is denoted as
$P_\mathrm{oi}$.  
In contrast to the traditional approach, where the top of the inner crust is determined 
by the neutron drip pressure at which neutrons drip out of nuclei 
(e.g., \citealt{Chamel_etal15_Drip}), in the nHD approach
$P_\mathrm{oi}$ 
is {\it not} 
uniquely defined 
by the ash composition and, generally, can not be predicted in advance.
In particular, it can vary
during FAC formation.
Generally, $P_\mathrm{oi}$ should be determined by the self-consistent calculations of the whole 
crust structure.

Another
basic feature of the nHD approach is a natural formation of the stationary self-similar FAC.
Accretion permanently delivers additional nuclei 
to
the crust
and,
to avoid accumulation of nuclei 
there
(i.e., to keep the total number of nuclei constant up to 
small secular changes, associated with the adjustment of hydrostatic structure of the crust to the increasing NS mass), 
a process of nuclei disintegration 
should take place.
This process naturally arises within the nHD approach in the form of nuclei disintegration 
instability 
deep inside 
the inner crust (see GC20, GC21, and upcoming \citealt{GC23_nHD_with_Shell} (GC23 in what 
follows) for 
more detailed analysis of the shell effects).
At the initial stage of accretion, when the pristine crust
is converted into FAC, the 
crust composition 
gradually changes,
in particular, the pressure $P_\mathrm{oi}$ at the oi interface 
gradually
varies with 
time. The total number of nuclei in the crust increases. This evolution 
is 
interrupted
by the onset of the instability: being activated, it burns out all additional nuclei, provided by 
subsequent accretion. 
In this way 
the FAC is formed; 
it has stationary structure and fixed $P_\mathrm{oi}$.

Theoretical determination of $P_\mathrm{oi}$ is a complicated problem. 
To find $P_{\rm oi}$, one should perform self-consistent calculations
of crust composition starting from the beginning of the accretion stage.
These calculations, obviously,  depend on the rather uncertain and 
model-dependent nuclear physics near $\sim 0.5 \rho_\mathrm{sat}$ 
($\rho_\mathrm{sat}\approx0.16$~fm$^{-3}$ is the
nuclear saturation density), where the instability occurs.% 
\footnote{For the smoothed CLDM, which neglects the shell effects, 
	$P_\mathrm{oi}$ determination is simplified
	because the whole inner crust structure can be determined 
	(in fact, by solving algebraic equations, see GC20), if $P_\mathrm{oi}$ is given. In 
	particular, there is 
	only one value of $P_\mathrm{oi}$, for which nuclei disintegration instability takes place and 
	the 
	crust
	connects to the core in a thermodynamically consistent way (GC20, GC21). 
	Shell effects make the problem more complicated
	leading to a range of admissible $P_{\rm oi}$ (see GC23 for details).
}

Hopefully, as discussed in GC20, GC21, SGC21, SGC22, one can construct FAC model in two steps. In
the first step a set of FAC `candidate' models, parametrized by $P_\mathrm{oi}$ should be 
calculated, while, in the second step, accurate value of $P_\mathrm{oi}$ should be specified on the 
basis of theoretical and/or observational constraints.
Here, as in SGC21 and SGC22, we follow this approach 
but proceed a bit further
by presenting details of the nuclear evolution in the outer regions of the inner crust.
As realistic values of 
$P_\mathrm{oi}$, we consider the range of pressures near the pressure at the oi interface for the 
cold 
catalyzed crust, $P_\mathrm{nd}^\mathrm{(cat)}$.
In accordance with the results of GC20 and GC21, it is a region, 
where the oi interface is expected 
for FAC. 

In SGC21 
we calculated
the nuclear evolution and heating profiles 
in the outer crust. For this purpose we used a simplified reaction network, based on the typical 
reaction timescales.
We performed calculations for the three different models of thermonuclear burning ashes, 
used
as initial compositions:
Superburst and Kepler ashes (most abundant elements are $^{56}\mathrm{Fe}$, $^{64}\mathrm{Ni}$), 
and Extreme rp ashes (dominant nuclide is $^{104}\mathrm{Pd}$). We considered
uniformly mixed matter, i.e.\ we did not take into account the effects of the matter separation 
during crystallization (\citealt{Horowitz+07,Medin_Cumming10,Mckinven_ea16,Caplan_etal18,Baiko22}).

In SGC21 we 
employed the
three theoretical mass tables: HFB24 
(\citealt{HFB24}), 
FRDM92 (\citealt{FRDM95}), and FRDM12 (\citealt{FRDM12}), supplemented by the most recent 
experimental results, summarized in AME20 (\citealt{ame20}). This allowed us to 
investigate the effects of nuclear physics theoretical uncertainties. 
The chosen models were used to 
obtain the composition at the bottom of the outer crust, parametrized by the pressure 
$P_\mathrm{oi}$.

In SGC22 we extended the results of SGC21 to the upper layers of the inner crust 
and presented 
the energy release calculations, leaving the details of nuclear evolution for the present paper. 
We considered the top region of the inner crust (up to the densities 
$\rho_\mathrm{dc}\approx2\times 
10^{12}$~g\,cm$^{-3}$), where the number density of free neutrons is relativity low and
theoretical atomic mass tables can be applied to calculate the nuclear evolution 
(\citealt{lau_ea18,Schatz_ea22}).
However, HFB24 and FRDM92  models appeared insufficient to describe the considered inner crust 
region due to the lack of data on the neutron-rich nuclei. 
Thus, in SGC22 and here we presented results only for FRDM12 model, 
which contains wide enough range of neutron-rich nuclei for our simulations. 
To calculate contribution of the free neutrons to the equation of state (chemical potential, 
partial pressure and energy density) we used BSk24 equation of state 
(\citealt{Goriely_ea_Bsk22-26}).

To account for nHD condition in the inner crust, we considered compression of a volume element 
attached 
to nuclei within the simplified reaction network.
The numerical procedure was based on an iterative increase of the total pressure and accompanying 
increase of the neutron chemical potential $\mu_n$ in accordance with the nHD condition. Note that 
in the traditional framework only the former is increased, which, generally, leads to violation of the nHD 
equilibrium.  
Following steps were applied 
for the analysis of
nuclear evolution at each iteration:
(1) We checked for all allowed reactions in the volume element. A reaction was treated as allowed 
if it reduced the thermodynamic potential $\Psi$ (GKC21). (2) We performed all allowed reactions 
by small chunks according to priority rules: (a) emission/capture of neutrons, (b) electron 
emission/capture plus emission/capture of neutrons, (c) pycnonuclear fusion. We did not 
constrained
the number of emitted/captured neutrons for the reaction types (a) and (b). The pycnonuclear 
reactions are allowed, if their rate $\tau_\mathrm{pyc}$ exceeds 
the compression rate $\tau_\mathrm{acc}$ (see \citealt{SC19_MNRAS} for details); the reaction rates 
were calculated 
according to \cite{Yakovlev_ea06}
assuming temperature $T=5\times10^8$~K in the crust.
The 
astrophysical factors were taken from \cite{Afanasjev_ea12}. 
(3) We compressed the volume element to the 
next iteration by increasing the pressure and $\mu_n$.

The initial composition at $P=P_\mathrm{oi}$
was taken to be equal to the composition at the respective pressure, obtained for the outer crust 
in our earlier paper SGC21; the neutron chemical potential at this point was taken to be equal to 
the neutron 
bare mass.

%%%%%%%%%%%%%%%%%%%%%%%%%%%%%%%%%%%%%%%%%%%%%%%%%%%%%%%%%%%%%%%%%%%%%%%%%%%%%%
\section{Results}\label{Sec:res}
%%%%%%%%%%%%%%%%%%%%%%%%%%%%%%%%%%%%%%%%%%%%%%%%%%%%%%%%%%%%%%%%%%%%%%%%%%%%%%

%%%%%%%%%%%%%%%%%%%%%%%%%%%%%%%%%%%%%%%%%%%%%%%%%%%%%%%%%%%%%%%%%%%%%%%%%%
\begin{figure*}
	\includegraphics[width=1.5\columnwidth]{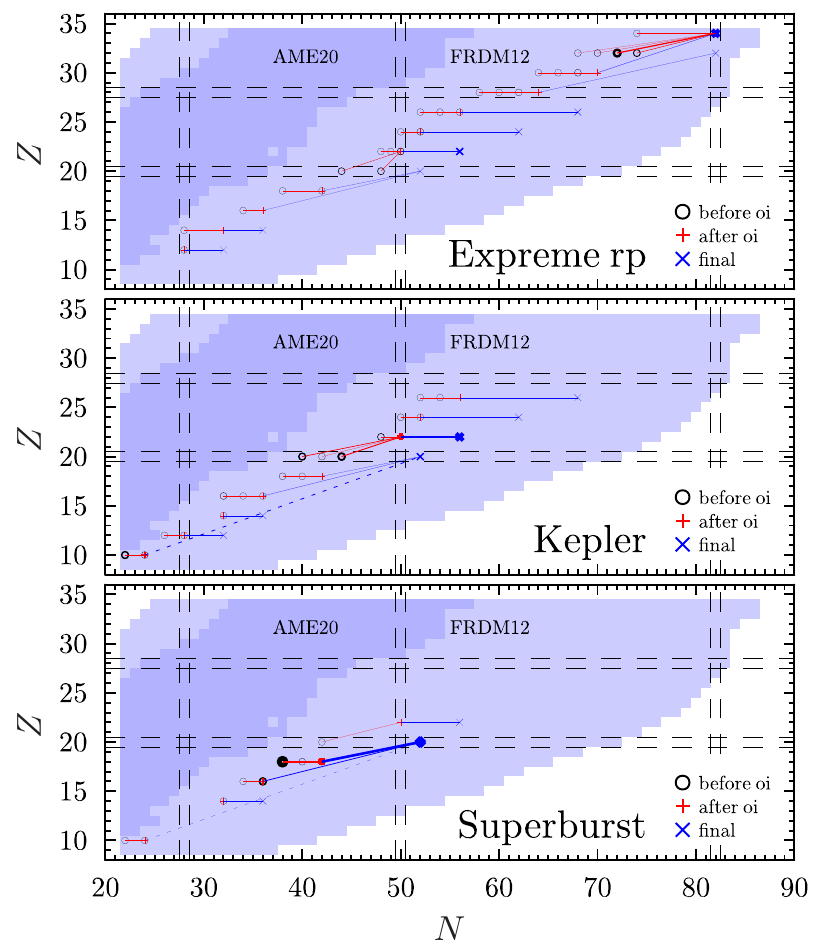}
	\caption{Schematic nuclear evolution in the upper layers of the inner crust for  
	$P_\mathrm{oi}=8\times 10^{29}$~dyn\,cm$^{-2}$. Black circles denote composition at the bottom 
	of the outer crust, red pluses mark composition after crossing the oi interface and blue 
	crosses the final composition in the end of simulation ($\rho=\rho_{\rm dc}$). 
	Red lines 
	describe net reaction path 
	at the oi interface, blue lines correspond to the net reaction in the upper layers of the inner 
	crust. Blue lines are dashed, if the fusion reaction take place in the inner crust. The nuclei 
	magic numbers are shown for convenience.
		Line and symbol thickness encodes abundance.
		Nuclei, whose mass is included in the AME20 data are shown by dark violet.
		Nuclei, available only in the FRDM12 data are shown by light violet.}
	\label{Fig_NZ}
\end{figure*}  
%%%%%%%%%%%%%%%%%%%%%%%%%%%%%%%%%%%%%%%%%%%%%%%%%%%%%%%%%%%%%%%%%%%%%%%%%%%

In this section we describe nuclear evolution of the three different thermonuclear ashes in the 
inner crust and its dependence on $P_\mathrm{oi}$. 
Let us start from 
discussing the
general trends.
As an illustrative example, we 
consider
the reaction flow for the crust models with 
$P_\mathrm{oi}=8.0\times 10^{29}$\,dyn\,cm$^{-2}$ ($\rho_\mathrm{oi}\approx 4.1\times 
10^{11}$~g\,cm$^{-3}$), shown  in Figure \ref{Fig_NZ}. 
In this figure we split nuclear evolution into two stages: reactions at the outer-inner crust 
interface and subsequent evolution 
up to
$\rho=\rho_\mathrm{dc}$.
Panels, from top to the bottom, correspond to 
Extreme rp, Kepler, and Superburst
ashes, respectively.
The composition at the bottom of the outer crust (i.e., at $P=P_\mathrm{oi}= 8\times 
10^{29}$~dyn\,cm$^{-2}$) is shown by black circles, composition after reactions at the oi interface 
is marked by the red pluses; finally,  blue crosses represent composition at the end of the 
simulation.
The red dotted lines 
show the net reactions (i.e., we do not show every single reaction) at the oi interface, blue lines 
represent net reactions inside the inner crust (dashes indicate that pycnonuclear fusion takes 
place). 
Thickness of symbols and lines encode abundances.

We demonstrate the reaction pathways on the mass chart layout, where the dark violet area 
corresponds 
to experimentally measured nuclei (AME20), while the light violet area corresponds to FRDM12 theoretical 
mass 
table. One can notice
that the nuclear evolution in the considered region lies well within the boundaries of the FRDM12 
table 
thus it is not directly relied on the experimental data and is not affected by the treatment of the tables 
boundaries.

As soon as nuclei enter the inner crust, they are immersed into the gas of free neutrons.
All energetically allowed neutron capture reactions start to proceed.
It is worth stressing that
neutron captures
do not deplete amount of free neutrons 
at a given layer, because the latter is set by the nHD 
condition. From the kinetic point of view, the captured neutrons are (quickly enough) replenished 
by 
neutron diffusion or superfluid flow.
As a consequence, nuclei  reach local equilibrium with respect to neutron captures/emissions at a 
given 
neutron chemical potential. Therefore, while in the outer crust several isotopes with the same $Z$ 
can 
exist, in the inner crust neutron captures/emissions drives all isotopes to the most favourable one 
for each $Z$.

Moreover, for nuclei crossing the oi interface from the outer crust, pure neutron captures 
(horizontal 
red lines in Figure \ref{Fig_NZ}) can be accompanied by electrons emissions (sloping red lines).
In particular, for Extreme rp  $^{100,102,104,106}\mathrm{Ge}$ transform to $^{116}\mathrm{Se}$ and 
$^{64,68}\mathrm{Ca}$ become $^{72}\mathrm{Ti}$. For Kepler ashes $^{60,62,64}\mathrm{Ca} 
\rightarrow {^{72}\mathrm{Ti}}$ reactions proceed.

It is important to highlight the role of the shell effects:
the reactions at the oi interface convert substantial part of nuclei to the shell closure 
$N=50, 82$.
They can also affect evolution in the shallow regions of the inner crust.
For example, for Extreme rp ashes $^{116}\mathrm{Se}$ nucleus ($N=82$) is formed at the oi 
interface and stay stable in the considered inner crust region 
(with some additional production of these isotopes from $^{100}\mathrm{Zn}$ nuclei).

For the applied FRDM12 mass table the ground state element 
at the oi interface is $^{118}\mathrm{Kr}$ and, generally, $Z\simeq 40$ corresponds the global 
minima through the upper layers of the catalyzed crust (e.g., \citealt{Pearson_ea18_bsk22-26}) and 
even 
accreted inner crust (GC23).
Since many nuclei at the bottom of the inner crust have $Z<40$ they tend to increase $Z$ by 
electron 
emissions. In some cases electron emission is  initiated by neutron captures.
However, $Z=40$  remains unreached by nuclei in our simulation.

Neutron captures and electron emissions remain the main reactions in the deeper layers of the inner 
crust.
In addition,
the thermally enhanced pycnonuclear fusion of light elements can proceed at high densities. In 
particular, this type of reactions burns out $Z=10$ nuclei, 
enriching the amount of $Z=20$ nuclei for 
Kepler and Superburst ashes (see dashed blue lines in Figure \ref{Fig_NZ}).

Summarizing, the main set of reactions is neutron captures, electron emissions and pycnonuclear 
fusion
(the latter noticeably 
affects crustal composition parameters, described in the next subsection, only for Kepler ashes, 
which have significant amount of light nuclei).
It is interesting, that
 in the traditional approach the inverse reactions, namely, electron captures and neutron 
emissions, were supposed to be dominating. 
This is due to the fact that neutron emission was the 
only source of unbound neutrons in the traditional models. 
On the opposite, within the nHD 
approach,
unbound neutrons in the shallow layers  of the inner crust 
are provided by diffusion/superfluid currents 
from the deeper regions.

%%%%%%%%%%%%%%%%%%%%%%%%%%%%%%%%%%%%%%%%%%%%%%%%%%%%%%%%%%%%%%%%%%%%%%%%%%%%%%
\subsection{Composition profiles for accreted inner crust}
%%%%%%%%%%%%%%%%%%%%%%%%%%%%%%%%%%%%%%%%%%%%%%%%%%%%%%%%%%%%%%%%%%%%%%%%%%%%%%

\begin{figure}
	\includegraphics[width=0.98\columnwidth]{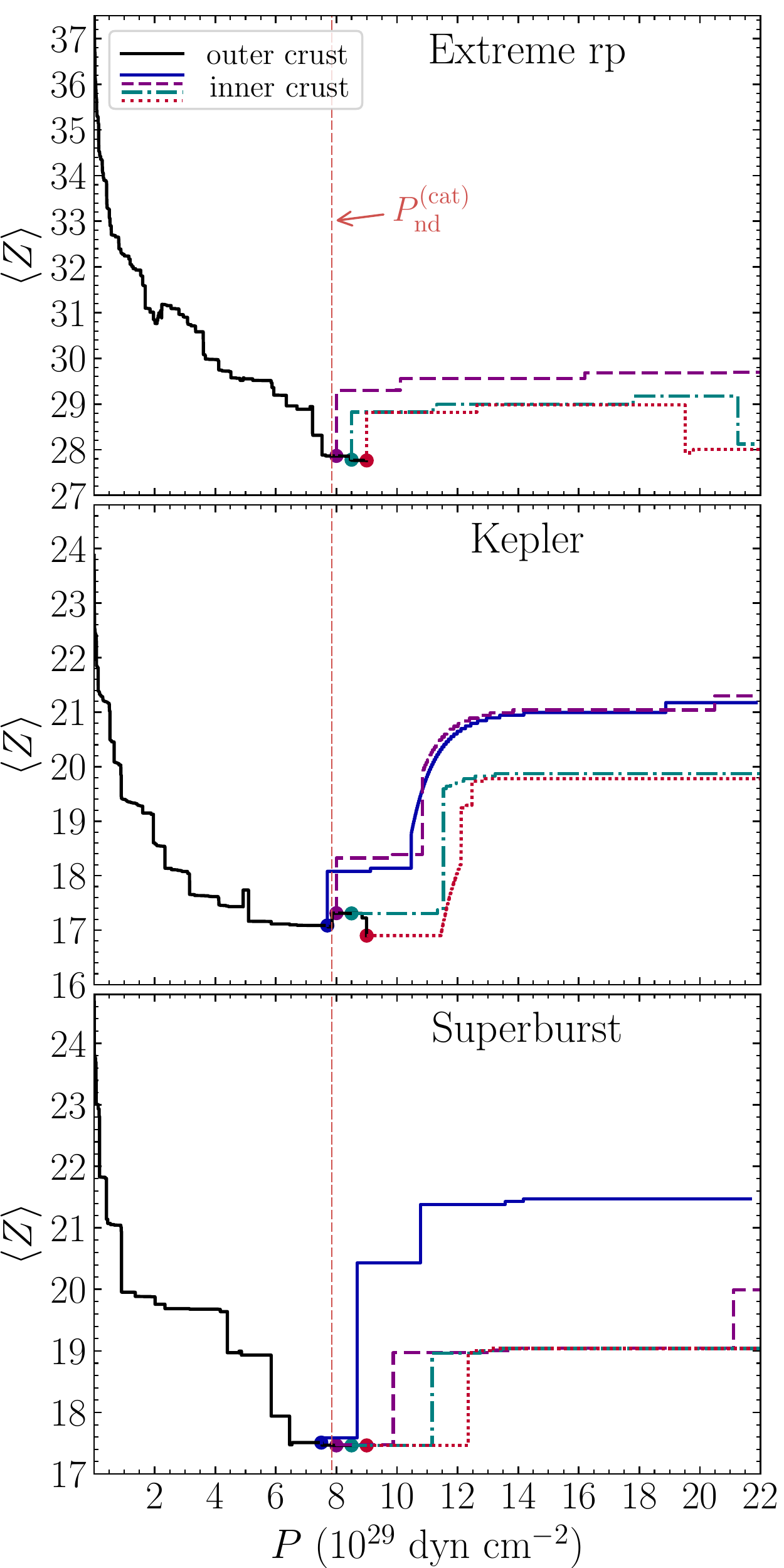}
	\caption{Profile of the average charge in the crust, $\langle Z\rangle$, versus pressure, $P$.  
		The three panels show the three different compositions of ashes considered in the paper. 
		The solid black lines correspond to the outer crust. Navy solid lines describe $\langle 
		Z\rangle$ evolution starting from $P_\mathrm{oi}=7.7\times 10^{29}$~dyn\,cm$^{-2}$ for 
		Kepler and $P_\mathrm{oi}=7.5\times 10^{29}$~dyn\,cm$^{-2}$ for Superburst ashes. Purple 
		dashed, green dash-dotted and red dotted lines trace the $\langle Z\rangle 
		(P_\mathrm{oi})$ dependence starting from $P_\mathrm{oi}=(8.0, 8.5, 9.0)\times 
		10^{29}$~dyn\,cm$^{-2}$, respectively. The oi interface is marked by circles. Vertical 
		dashed lines indicate $P_\mathrm{nd}^\mathrm{(cat)}$, the pressure at the oi interface 
		for the catalyzed crust.}
	\label{Fig_Z}
\end{figure}

\begin{figure}
	\includegraphics[width=0.98\columnwidth]{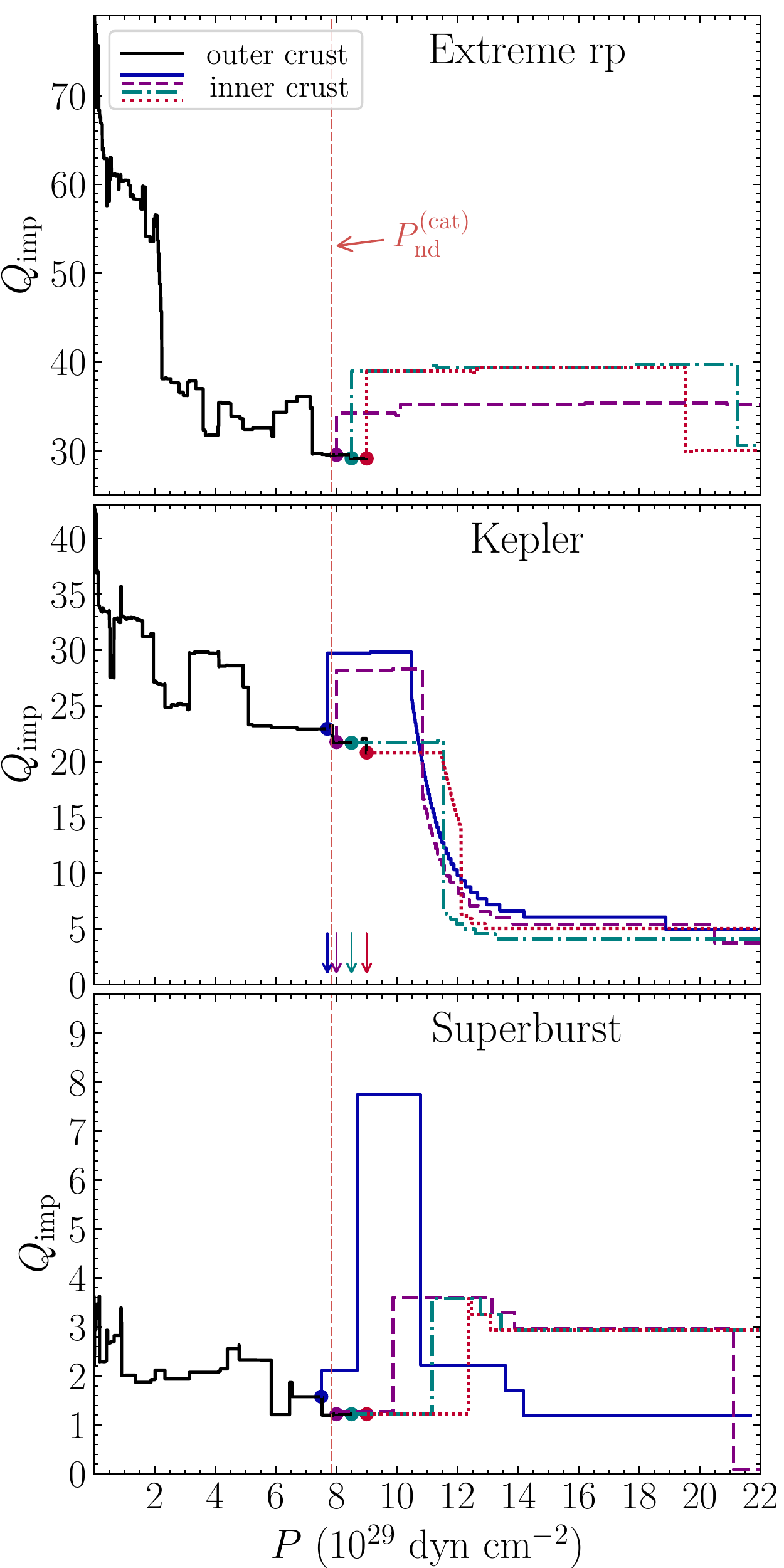}
	\caption{$Q_\mathrm{imp}$ profile in the crust. The legend is the same as in Figure 
	\ref{Fig_Z}. 
	Colored arrows mark the $P_\mathrm{oi}$ for Kepler ashes.}
	\label{Fig_Qimp}
\end{figure}

In Figures~\ref{Fig_Z}, \ref{Fig_Qimp}  we demonstrate  crustal composition parameters $\langle 
Z\rangle=\sum_{i}X_i Z_i$ and $Q_\mathrm{imp}=\sum_{i}X_i(Z_i-\langle Z\rangle)^2$ as functions of 
pressure for a set of models, parametrized  by the pressure $P_\mathrm{oi}$ at the outer-inner 
crust interface 
(here 
$Z_i$ and $X_i$ denote charge and fractional number of nuclei of type $i$). 
Namely, we choose the 
representative set of $P_\mathrm{oi}= (8, 8.5, 9)\times 10^{29}$~dyn\,cm$^{-2}$. 
In addition, we 
add  $P_\mathrm{oi}= 7.7\times 10^{29}$~dyn\,cm$^{-2}$ for Kepler ashes and $P_\mathrm{oi}= 
7.5\times 10^{29}$~dyn\,cm$^{-2}$ for Superburst ashes as a reference lower bounds on 
$P_\mathrm{oi}$, which were obtained previously (see SGC22). For the Extreme rp ashes the lower 
limit 
is $P_\mathrm{oi}= 8\times 10^{29}$~dyn\,cm$^{-2}$, and is already included into the set.

As discussed above, electron emissions
often occur
at oi interface.
They increase the average charge $\langle Z\rangle$ and impurity parameter $Q_\mathrm{imp}$ in all 
the presented cases for Extreme rp ashes and for the two of Kepler models with low $P_\mathrm{oi}$ 
(see Figures~\ref{Fig_Z} and \ref{Fig_Qimp}). Meanwhile, for the Superburst ashes electron 
emissions are 
insignificant at the oi interface.

In the inner crust, the neutron chemical potential increases, leading to additional neutron
captures
stimulating electron emissions, if the latter are not blocked by the growth of $\mu_\mathrm{e}$. 
These 
reactions act for all the three ashes 
so that
$\langle Z\rangle$ grows in the shallow inner crust 
region 
for all models. For Kepler ashes the additional 
source of 
$\langle Z\rangle$ growth is pycnonuclear fusions.
However, for 
models with 
high $P_\mathrm{oi}$ and 
Extreme rp ashes,
$\langle Z\rangle$ is reduced near 
the end of our simulation 
(the relevant reactions are discussed below).

The behaviour of the impurity parameter is not that simple.  
Let us discuss it along with the underlying reactions for each of the 
considered thermonuclear ash compositions.
%
%%%%%%%%%%%%%%%%%%%%%%%%%%%%%%%%%%%%%%%%%%%%%%%%%%%%%%%%%%%%%%%%%%%%%%%%%%%%%%%%%%%%%%%%%%%% 


For Extreme rp ashes   $Q_\mathrm{imp}$ and $\langle Z \rangle$ jump at the oi interface  for all 
the considered $P_\mathrm{oi}$ cases. 
A substantial contribution to these jumps
is provided by  
the reactions $^{102,104,106}\mathrm{Ge} 
\rightarrow {^{116}\mathrm{Se}}$ and $^{64,68}\mathrm{Ca} \rightarrow {^{72}\mathrm{Ti}}$.
However, jump in $Q_\mathrm{imp}$ 
is 
smaller for $P_\mathrm{oi}= 8.0\times 10^{29}$~dyn\,cm$^{-2}$, than 
for $P_\mathrm{oi}= (8.5, 9.0)\times 10^{29}$~dyn\,cm$^{-2}$.
This is because the latter models have larger $\mu_\mathrm{e}$
at the oi interface,
which
blocks the reaction $^{64,68}\mathrm{Ca} \rightarrow {^{72}\mathrm{Ti}}$.
Suppression of this reaction also leads to lower $\langle Z\rangle$
for $P_\mathrm{oi}= (8.5, 9.0)\times 10^{29}$~dyn\,cm$^{-2}$.

The nuclear evolution 
deeper in 
the inner crust is affected by the compositional changes 
at the 
top of the inner crust, as well as by
the value of 
$P_\mathrm{oi}$
itself.
Namely, the higher $P_\mathrm{oi}$ 
the higher is $\mu_\mathrm{e}$ at the same pressure (the neutron
partial pressure is lower due to 
lower $\mu_\mathrm n$).
For the considered examples this leads to electron captures accompanied by neutron emissions
for models with $P_\mathrm{oi}= (8.5, 9.0)\times 10^{29}$~dyn\,cm$^{-2}$ near the end of calculations.
Such reactions return $^{116}\mathrm{Se}$ to $^{114}\mathrm{Ge}$, i.e. the net reaction for 
$\mathrm{Ge}$ isotopes, coming from the outer crust, becomes capture of neutrons, which does not 
change the charge number. As a result,  for models with $P_\mathrm{oi}= (8.5, 9.0)\times 
10^{29}$~dyn\,cm$^{-2}$
we obtain almost the same  
$Q_\mathrm{imp}$ and $\langle Z \rangle$ 
in the end of our simulation (at $\rho=\rho_{\rm dc}$),
as at the bottom of the outer crust.

The situation is more complicated for 
models based on the
Kepler ashes.
The first two models (navy solid and purple dashed lines, $P_\mathrm{oi}= (7.7, 8.0)\times 
10^{29}$~dyn\,cm$^{-2}$) slightly differ by the composition at the bottom of the outer crust, 
although 
this difference only barely influence the behaviour in the inner crust. 
For these two models the main reactions at the oi interface 
are transformations of calcium isotopes into titanium: 
$^{60,62,64}\mathrm{Ca} \rightarrow {^{72}\mathrm{Ti}}$. 
At  $P\approx 1.0\times 10^{30}$~dyn\,cm$^{-2}$, $^{34}\mathrm{Ne}$ captures neutrons to become  
$^{38}\mathrm{Ne}$, whose typical pycnonuclear fusion time, $\tau_\mathrm{pyc}$, is 
approximately ten times shorter than for $^{34}\mathrm{Ne}$ due to a larger nucleus size, which 
reduces the Coulomb barrier and
simplifies tunneling.
This is clearly 
seen
in the astrophysical factors, 
differing
by
several orders of magnitude 
for various isotopes (see, e.g., \citealt{Beard+10,Afanasjev_ea12,Singh+19}). 
Within our simplified reaction network all $^{34}\mathrm{Ne}$  nuclei are converted into 
$^{38}\mathrm{Ne}$ 
in one and the same crustal layer, where  $\mu_n$ reaches the threshold for this reaction.
Pycnonuclear fusion $^{38}\mathrm{Ne}+^{38}\mathrm{Ne}$ occurs at the same pressure, 
and the
subsequent
neutron 
emissions
by the compound nuclei 
lead to formation of ${^{70}\mathrm{Ca}}$.
However, the fusion is not complete, 
leaving substantial amount of $^{38}\mathrm{Ne}$ in the mixture.
This is because fusion 
reduces the number of 
$^{38}\mathrm{Ne}$ 
nuclei
and hence pycnonuclear reaction 
rate.
According to our algorithm,  
this reaction stops operating
just after the reaction timescale 
$\tau_\mathrm{pyc}$  falls down to the value $\tau_\mathrm{acc}$. 
Subsequent burnout of $^{38}\mathrm{Ne}$ 
occurs gradually,
being controlled by the condition $\tau_\mathrm{pyc} \approx \tau_\mathrm{acc}$.%
%
%%%%%%%%%%%
\footnote{\label{Footnote_SimpMod}
	Our simplified reaction network 
	will likely affect the details of this evolution, but we expect 
	the 
	related artificial effects are not crucial, i.e. 
	accurate reaction network should lead to the same 
	net reaction: conversion of $^{34}\mathrm{Ne}$ into ${^{70}\mathrm{Ca}}$ in the more or less 
	the 
	same pressure region. 
	Note that the pycnonuclear reaction rate is very uncertain and state-of-the-art models can 
	differ 
	by orders of magnitude (see, e.g., \citealt{Yakovlev_ea06}). Thus, predictions of the 
	detailed 
	reaction network would in any case be sensitive to the applied model. 
	In particular, if the maximum reaction rate model of \citealt{Yakovlev_ea06} was applied, it 
	would 
	lead to the complete pycnonuclear burning of $^{34}\mathrm{Ne}$ already in the outer crust.}
%%%%%%%%
%
This behaviour is displayed in the Figures~\ref{Fig_Z} and \ref{Fig_Qimp}, where $\langle Z\rangle$ 
gradually increases while $Q_\mathrm{imp}$  decreases, approaching nearly constant value.
The difference
between $P_\mathrm{oi}= 7.7\times 10^{29}$~dyn\,cm$^{-2}$
and $P_\mathrm{oi}= 8.0\times 10^{29}$~dyn\,cm$^{-2}$ models is associated with the fact that for 
the second model the neutron captures occur at higher pressure, leading to higher number density of 
$^{38}\mathrm{Ne}$ and larger fraction of these nuclei burned just after formation. 

The main difference of two remaining  high $P_\mathrm{oi}$ models for Kepler ashes  is the 
suppression of electron emissions at the oi interface. 
The further details of reaction pathways also 
differ from two previously discussed models, however, the most important net reaction for 
the subsequent evolution is the same as in previous models: $^{34}\mathrm{Ne} +^{34}\mathrm{Ne} 
\rightarrow 
{^{70}\mathrm{Ca}}$. For the model with $P_\mathrm{oi}= 8.5\times 10^{29}$~dyn\,cm$^{-2}$ (green 
dash-dotted line) 
$^{38}\mathrm{Ne}$ almost completely fuses to ${^{70}\mathrm{Ca}}$, 
just after $^{38}\mathrm{Ne}$ is formed by neutron captures
(see the big vertical 
drop
of the  line).
However for the last model with the highest $P_\mathrm{oi}$ 
the shift of oi interface leads to the 
increase
of the threshold pressure for neutron captures by 
$^{34}\mathrm{Ne}$ 
due to a lower neutron number density at a given pressure.
As a result, within our simplified reaction network, pycnonuclear fusion occurs already with 
$^{34}\mathrm{Ne}$ but, as for the models with lower $P_\mathrm{oi}$, it is not complete. Only 
about a half 
of $^{34}\mathrm{Ne}$ is burned, while the remaining part captures neutrons and rapidly burns  
after that. This is 
illustrated
by 
the red dotted line in Figure~\ref{Fig_Z}, where $\langle 
Z\rangle$ gradually grows at 
$P\approx 1.2\times 10^{30}$~dyn\,cm$^{-2}$ 
and then 
exhibits a
jump, associated with the  neutron capture threshold, leading to
$^{38}\mathrm{Ne}$ formation and rapid  pycnonuclear fusion. 
The second difference of the last model is directly  associated with  composition at the bottom of 
the outer crust. Specifically, the lightest calcium isotopes are already transformed to 
$^{60}\mathrm{Ar}$ in the outer crust, and this 
results in 
the lower $\langle Z\rangle$ value at the oi 
interface. 
However, these nuclides are finally converted into $^{70}\mathrm{Ca}$ in almost the same region, 
where pycnonuclear fusion of $\mathrm{Ne}$ takes place. As a result,  $Q_\mathrm{imp}$ and $\langle 
Z\rangle$ in the end of our simulation is almost identical for the last two models.
$Q_\mathrm{imp}$ and $\langle Z\rangle$ are not affected by neutron captures
deep inside the 
inner crust and the most abundant element in the end of simulation for two high-$P_\mathrm{oi}$ 
models 
is 
$^{72}\mathrm{Ca}$,
while for two low-$P_\mathrm{oi}$ models it is
${^{78}\mathrm{Ti}}$.
This is the main reason for the difference in the final $\langle Z\rangle$ between the two sets of 
models.

Finally, for the Superburst ashes, the major trends are as follows.
The shift of the oi interface changes the initial composition and affects the subsequent evolution. 
Namely, in the first 
model  
(with $P_\mathrm{oi}= 7.5\times 10^{29}$~dyn\,cm$^{-2}$)
$^{60}\mathrm{Ca}$  
transforms into $^{72}\mathrm{Ti}$ at the oi interface  (see the modest initial raise of the blue 
curve in Figures~\ref{Fig_Z} and \ref{Fig_Qimp}). Meanwhile, for models with larger $P_\mathrm{oi}$,
$^{60}\mathrm{Ca}$ is already converted into $^{60}\mathrm{Ar}$ in the outer crust, replenishing the amount of the latter in contrast to the first model, and no important reactions occur at the oi interface.
$^{60}\mathrm{Ar}$, which  is contained in all models, is
transformed into heavier nuclei in the depths of the inner crust, causing the 
second jump for the model with $P_\mathrm{oi}= 7.5\times 10^{29}$~dyn\,cm$^{-2}$ and 
consecutive first jumps for higher $P_\mathrm{oi}$ models in $\langle Z\rangle$ and 
$Q_\mathrm{imp}$ in Figures~\ref{Fig_Z} and \ref{Fig_Qimp} 
(note that the pressure for such transitions increases with $P_\mathrm{oi}$).
For the three high $P_\mathrm{oi}$ models  $^{60}\mathrm{Ar}$ is converted into $^{70}\mathrm{Ca}$, 
however, for the 
lowest
$P_\mathrm{oi}$ model, 
the reaction  $^{70}\mathrm{Ca}\,\rightarrow$ $^{72}\mathrm{Ti}$, triggered by electron emission 
plus neutron capture also 
proceeds, increasing the abundance of $^{72}\mathrm{Ti}$ formed at the oi interface and leading to a much more
pronounced
peak in $Q_\mathrm{imp}$ (see Figure 
\ref{Fig_Qimp}, note the scale of the Superburst panel). Then 
the jump of $\langle Z \rangle $ and decrease of $Q_\mathrm{imp}$ for the first model at $P\approx 
1.1\times 10^{30}$~dyn\,cm$^{-2}$ is associated with the reaction $^{54}\mathrm{S} \rightarrow 
{^{70}\mathrm{Ca}}$. A similar reaction occurs for the second model near the end of the simulation 
$^{58}\mathrm{S} \rightarrow {^{72}\mathrm{Ca}}$ (see the growth of $\langle Z\rangle$ and fall of 
$Q_\mathrm{imp}$ for the dashed purple curve), but for other models
$^{58}\mathrm{S}$
survives up to the end of our simulations.
The pycnonuclear fusion is less important for Superburst ashes in comparison to Kepler ashes due 
to lower abundance of light isotopes.
However, it takes place at 
$P\approx (13-14)\times 10^{29}$~dyn\,cm$^{-2}$ 
and leads to 
the same net reaction 
$^{38}\mathrm{Ne}+^{38}\mathrm{Ne} \rightarrow {^{70}\mathrm{Ca}}$, 
triggered by compression.
The fusion occurs a bit earlier for higher $P_\mathrm{oi}$ models due to smaller amount of free 
neutrons and higher nuclei concentration (at a given pressure).
In Figure~\ref{Fig_Qimp} the fusion reveals itself by two modest 
drops
of $Q_\mathrm{imp}$ at sufficiently large $P$.

%%%%%%%%%%%%%%%%%%%%%%%%%%%%%%%%%%%%%%%%%%%%%%%%%%%%%%%%%%%%%%%%%%%%%%%%%%%%%%%%%%%%%%%%%%%% 
\subsection{Equation of state}
%%%%%%%%%%%%%%%%%%%%%%%%%%%%%%%%%%%%%%%%%%%%%%%%%%%%%%%%%%%%%%%%%%%%%%%%%%%%%%%%%%%%%%%%%%%% 

\begin{figure}
	\includegraphics[width=0.992\columnwidth]{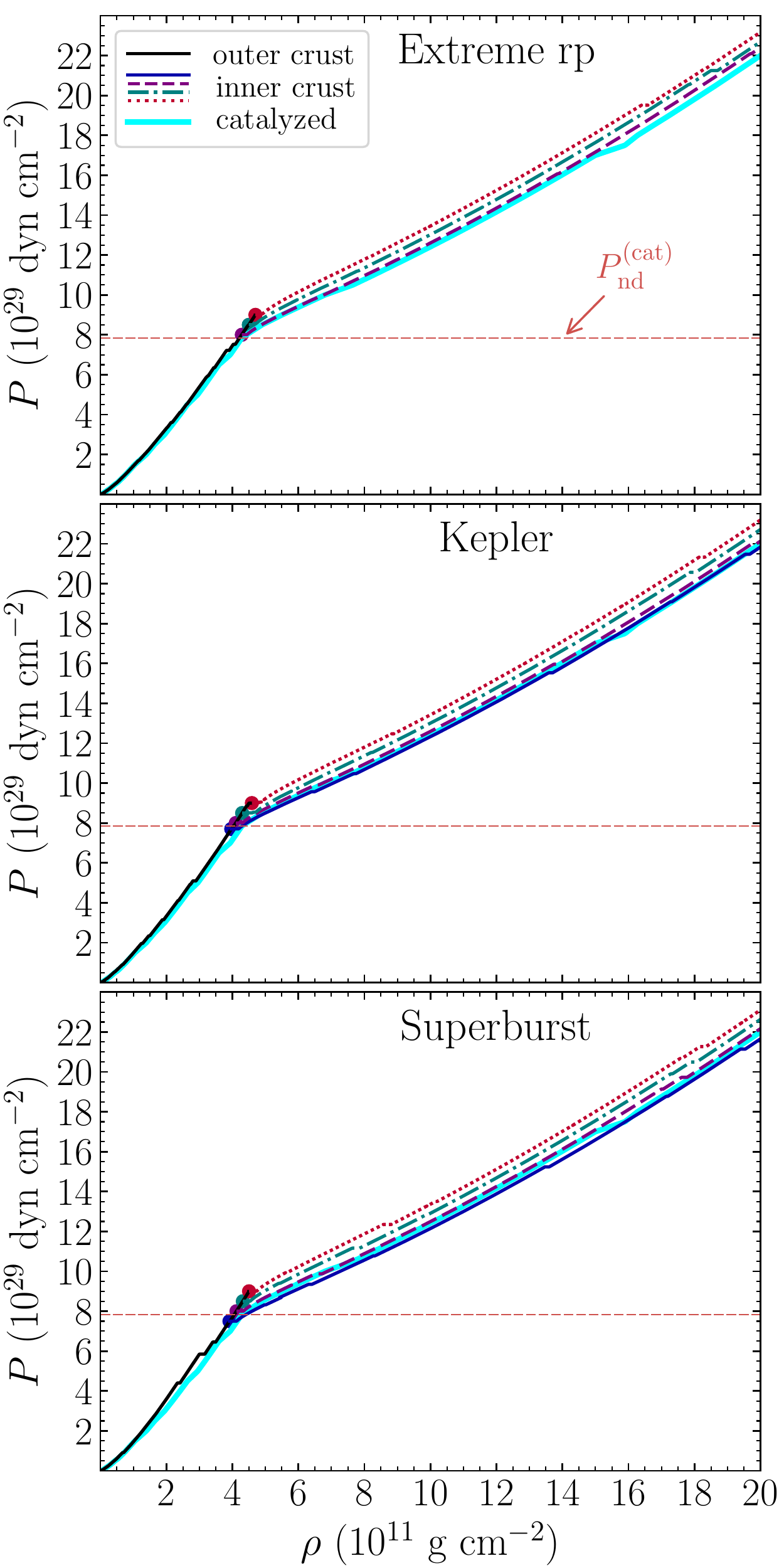}
	\caption{Equation of state in the crust for Extreme rp, Kepler, and Superburst ashes and 
	different $P_{\rm oi}$.
	The accreted crust models are 
	%indicated 
	the same
	as in Figure \ref{Fig_Z}.  For comparison,  the catalyzed EOS is shown by  the solid cyan 
	line.} 
	\label{ro}
\end{figure}

In Figure \ref{ro} we demonstrate the accreted crust equation of state (EOS) for the three ash 
compositions 
(Extreme rp, 
Kepler, and Superburst) with different 
locations of the outer-inner crust interface (marked by circles). 
In addition,
we added catalyzed EOS (cyan line), calculated with the same nuclear physics input.
All accreted crust curves show similar behaviour, which generally follow the catalyzed crust 
EOS. In particular, the oi transition is accompanied by the softening of the EOS, corresponding to 
the appearance of unbound neutrons. 
In contrast,
the accreted crust EOS in the traditional 
approach (see, e.g. \citealt{HZ90b,Fantina+22})
does not demonstrate significant softening after the oi interface, in particular, the 
adiabatic index stays larger than 1.4 (see Figures 1 and 2 in \citealt{Fantina+22}). This
leads to the overestimation of NS radii in the traditional models.

It should be mentioned, that in our approach reactions are generally accompanied by the density 
jumps. This stems from the fact, that 
nucleon
number in the nuclei changes by integer value 
(usually this is an even number due to pairing correlations).
Therefore,
in a
general reaction both $\langle N\rangle$ and $\langle Z\rangle$ 
 experience a jump. 
 At the same time,  
the free neutron number density stays constant, 
while 
the electron number density 
readjusts only slightly in order
to keep the pressure fixed 
(the readjustment is needed to balance small Coulomb terms, 
which are affected by the change of $\langle Z\rangle$).
The density usually increases after reaction.
However, in some cases the density may decrease.
All these cases are closely related to 
neutron shell effects, which strongly 
influence energies of nuclei for the applied FRMD12 mass 
model.
Note, however, that 
the reality of these effects are somewhat questionable,
because the realistic band-structure calculations by \cite{Chamel05, Chamel06, Chamel12} indicate 
that unbound neutron states form a quasi-continuum.
Moreover, most of the density drops 
found by us
are
artefacts of our simplified reaction network, 
which allows for small, but still finite portions of matter to react at once. 
Specifically,
for the pycnonuclear fusions, triggered by 
compression, a discrete amount of $^{38}\mathrm{Ne}$ transforms into ${^{70}\mathrm{Ca}}$, where 
the 
shell closure $N=50$ is occupied. This leads to the modest increase of $Y_\mathrm{p}=\langle 
Z\rangle /\langle N+Z\rangle$ and, as a consequence, 
to the density decrease.
In our calculations, the pycnonuclear reaction converts two chunks of $^{38}\mathrm{Ne}$ into 
${^{70}\mathrm{Ca}}$, leading to a density drop, but only by about $0.01\%$
of the total density. 
After that the abundance of fuel nuclei 
decreases, leading to reduction of the pycnonuclear reaction rate, which becomes slower  than the compression rate. 
In our simplified reaction network 
the pycnonuclear burning of the next chunks occurs only after some additional compression of 
matter, so that, in total, the density increases with the growing pressure in the pycnonuclear 
burning region. Therefore, we expect that the observed small drops will 
disappear 
in a more detailed reaction network, leading to a monotonic increase  of the density.
Still, 
for the Superburst ashes there is one transition, where the density falls by $\lesssim 1\%$
due to conversion of abundant isotope $^{60}\mathrm{Ar}$ into $^{72}\mathrm{Ti}$ (again, 
$N=50$) via the electron emissions and neutron captures.	
If the neutron star crust was a liquid, this would lead to the Rayleigh-Taylor instability
and mixing of matter in the vicinity of this reaction threshold. 
This possibility is indeed intriguing, but 
we 
avoid speculating on this subject further.
The main reason is the fact that the crust is likely 
crystallized 
near 
this 
transition 
(the 
melting temperature in this region $T_\mathrm{m}\approx10^9$\,K is larger than typical temperatures 
in the crust of accreting neutron stars).
In this case, according to \cite{Blaes_ea90}, the crust elasticity should suppress  the  
Rayleigh-Taylor instability for a $1\%$ density jump. This justifies our approach, which do not 
allow for
mixing of nuclei from adjacent layers.

%%%%%%%%%%%%%%%%%%%%%%%%%%%%%%%%%%%%%%%%%%%%%%%%%%%%%%%%%%%%%%%%%%%%%%%%%%%%%%%%%%%%%%%%%%%%
\section{Summary}\label{Sec:sum}
%%%%%%%%%%%%%%%%%%%%%%%%%%%%%%%%%%%%%%%%%%%%%%%%%%%%%%%%%%%%%%%%%%%%%%%%%%%%%%%%%%%%%%%%%%%% 
In this paper we 
discuss
nuclear reaction pathways and the composition profiles for nHD 
accreted crust models, 
starting from realistic composition of thermonuclear ashes (Extreme rp, 
Kepler and Superburst ashes). 
The heat release profiles for these models have already been presented in SGC22.

We implemented a multi-component simplified reaction network, which takes into account the nHD 
condition.
The developed network allows us to 
determine the reaction pathways
in the crust up to the density $\rho\approx 2\times 
10^{12}$~g\,cm$^{-3}$ (thus extending the results of SGC21, where the pathways were obtained only 
for 
the 
outer crust).
By varying the pressure $P_\mathrm{oi}$ at the outer-inner crust interface,
we generate a set of models, parametrized by the pressure $P_\mathrm{oi}$
which was considered as a free parameter of the theory.
In doing that, we
employed
the FRDM12 mass table 
supplemented with 
the BSk24 equation of state 
for unbound neutrons (see section \ref{Sec:Approach} for details).

In our simulations
neutron captures and electron emissions are shown to be the most important reactions, which govern 
nuclear 
evolution and energy release in the inner crust.
This is in sharp contrast with the traditional approach, in which inverse reactions mainly take 
place 
(see, e.g., 
\citealt{HZ08, lau_ea18}). 
The pycnonuclear fusion, which was supposed to be crucial 
in the
traditional approach, 
plays some role
only for Kepler ashes. 
We find no pycnonuclear fusion cycles (see \citealt{lau_ea18, SC19_MNRAS} for a discussion of 
these 
cycles within the traditional approach). 

The main output of our calculations is the 
profiles of $\langle Z\rangle$ and $Q_\mathrm{imp}$ as  functions of 
pressure for a number
of models, 
parametrized by 
different $P_\mathrm{oi}$. 
Namely, the considered set of $P_\mathrm{oi}$ 
includes a lower limit on this parameter, obtained in \cite{SGC22a}, as well as 
$P_\mathrm{oi}=(8.0, 
8.5, 9.0)\times 10^{29}$~dyn\,cm$^{-2}$. 
In addition, we present equation of state for all the 
models.

We found $\langle Z\rangle$ to be non-monotonic function of pressure: after an almost monotonic 
decrease in the outer crust, $\langle Z\rangle$ passes a local minimum at the oi interface, then 
grows and approaches an almost constant value by the end of the simulation ($\langle 
Z\rangle\approx 19-21$ for Kepler and Superburst ashes and $\langle Z\rangle\approx 28-30$ for 
Extreme rp ashes). This behavior 
differs from 
the predictions of the traditional approach (e.g.,  
\citealt{HZ08, Fantina_ea18}), where $\langle Z\rangle$ generally decreases  after oi interface and 
increases only due to pycnonuclear reactions.\footnote{Note that the location of the oi interface 
	in 
	traditional models is placed deeper in the crust.} 
Typically (but not always),  $\langle Z\rangle$ profile lies lower for models with higher 
$P_\mathrm{oi}$.

In turn, $Q_\mathrm{imp}$ profiles strongly depend on the thermonuclear ash composition (see 
section 
\ref{Sec:res}) and, in the inner crust, differ from the results, obtained within traditional 
appoach (e.g., \citealt{lau_ea18, Schatz_ea22}). 
For Extreme rp ashes $Q_\mathrm{imp}$ 
reduces substantially
in the outer crust,
slightly increases
at the oi interface 
and stays
around
$30-35$  in the upper layers of the inner 
crust. 
In contrast, for Kepler ashes 
$Q_\mathrm{imp}$ 
decreases
from $\approx 45$ to $\approx 20$ in
the outer crust, 
then  pycnonuclear fusion in the inner crust further purifies the composition, pushing 
$Q_\mathrm{imp}$ down to $\approx 5$. 
For the Superburst ashes  $Q_\mathrm{imp}\approx 1-4$; only for the model with the lowest 
$P_\mathrm{oi}= 7.5\times 10^{29}$~dyn\,cm$^{-2}$ there is a rather narrow region with 
$Q_\mathrm{imp}\approx 8$.

As in the traditional approach (e.g., \citealt{lau_ea18,SC19_MNRAS}) the shell effects appear to be 
crucial
for the accreted crust composition.
The dominating role of
neutron capture reactions for nHD approach 
leads to occupation of $N=50, 
82$ neutron shell closures just after the oi interface (see figure~\ref{Fig_NZ} for Kepler and 
Extreme 
rp ashes). It is mainly the shell effects 
that 
are responsible for a complicated inner crust composition, 
spanning for several energy minima 
(e.g. 
for 
Extreme rp and Kepler ashes) and leading to 
large $Q_\mathrm{imp}$. 
Note that for Superburst ashes, almost all 
nuclei transform into the 
nuclide with $Z=20$. 

We also determined EOS in the accreted crust (for $\rho <\rho_{\rm dc}$) for all the three ash compositions and models with different $P_\mathrm{oi}$.
The $P(\rho)$ curves for accreted crust lie close to each other, being quite similar to the catalyzed EOS. Specifically, one can notice a softening of the EOS at the oi interface, associated with the appearance of free neutrons.
Note, that in the traditional approach the softening does not occur (\citealt{HZ90b, Fantina+22}).
Another important feature is the density jumps, produced by nuclear reactions 
in our simplified reaction network.
These jumps occur because the number of nucleons
in
nuclei changes by an integer.
The jumps are also affected by pairing correlations and shell effects. 
We find that reactions can  
slightly decrease the density, but at most by $1\%$, 
which 
is
not sufficient
for triggering the Rayleigh-Taylor instability in the solidified crust (\citealt{Blaes_ea90}).
It should be noted that the jumps 
will likely be
smoothed out 
if a more realistic reaction network is applied.

Our results for the composition profile 
and
the calculated heating (\citealt{SGC22a}) can be 
implemented for modeling the neutron star cooling during quiescence periods. 
By confronting the
observational data with cooling models one may try to
constrain the key parameter of our theory, the
pressure 
$P_\mathrm{oi}$, 
along with 
other microscopic parameters of neutron star crust.
Note,  that all up-to-date cooling models available in the literature (e.g., 
\citealt{syhp07,bc09,pr12,Deibel+15,wdp17,Brown_ea18,mdkse18,pc21,Parikh_ea21,Page+22,Mendes_ea22}) 
are based on the results of the traditional approach.
This means that the constraints on the properties of superdense matter 
obtained from these models
should be revisited,
and this can be done
by employing the nHD models of the accreted crust developed in the present paper.

%\vspace{-0.5cm}
%%%%%%%%%%%%%%%%%%%%%%%%%%%%%%%%%%%%%%%%%%%%%%%%%%%%%%%
\section*{Acknowledgements}
%%%%%%%%%%%%%%%%%%%%%%%%%%%%%%%%%%%%%%%%%%%%%%%%%%%%%%%
The work of N.~N.\ Shchechilin (calculations and original draft preparation) was funded by the FWO (Belgium) and the F.R.S.-FNRS (Belgium) under the Excellence of Science (EOS) programme (project No. 40007501).
The work of M.~E.\ Gusakov (theoretical framework and development of an independent code verifying 
the numerical results obtained in the paper) and A.~I.\ Chugunov (theoretical framework) was 
supported by Russian Science Foundation (grant 22-12-00048).
%\vspace{-0.5cm}
%%%%%%%%%%%%%%%%%%
\section*{DATA AVAILABILITY}
%%%%%%%%%%%%%%%%%%%%%%%%%%%%%%%%%%%%%%%%%%%%%%%%%%%%%%%
Data can be provided by the authors upon a reasonable request.

%\vspace{-0.5cm}
\bibliographystyle{mnras}
\bibliography{literature}

\label{lastpage}

\end{document}